\documentstyle[11pt,aaspp4]{article}

\def\EeV{\,{\rm EeV}}

\def\yr{\,{\rm yr}}

\def\kpc{\,{\rm kpc}}
\def\Mpc{\,{\rm Mpc}}

\def\cmm2{{\,\rm cm^{-2}}}
\def\cm2{{\,{\rm cm}^2}}
\def\cmm3{{\,{\rm cm}^{-3}}}
\def\gcmm3{{\,{\rm g\,cm^{-3}}}}

\def\G{\,{\rm G}}
\received{RECEIPT DATE}
\revised{REVISION DATE}
\accepted{ACCEPT DATE}
\cpright{AAS}{1996}

\journalid{VOL}{JOURNAL DATE}
\articleid{START PAGE}{END PAGE}
\paperid{MANUSCRIPT ID}
\cpright{AAS}{1995}
\ccc{CODE}

\lefthead{Lemoine {\it et al.}}
\righthead{Extragalactic Magnetic Field and Ultra-High Energy
Cosmic Rays}

\begin{document}
\vspace*{1cm}
\rightline{astro-ph/9704203}
\rightline{Submitted to \it Astrophysical Journal Letters}
\vspace{1cm}
\title{ULTRA-HIGH ENERGY COSMIC RAY SOURCES AND LARGE SCALE
MAGNETIC FIELDS}
\author{Martin Lemoine,  G\"unter Sigl, Angela V. Olinto, \& 
David N. Schramm}
\affil{Department of Astronomy \& Astrophysics\\
Enrico Fermi Institute, The University of Chicago, Chicago, IL
60637-1433\\}

\begin{abstract}

Protons of energies up to $\sim 10^{20}\,$eV can be subject to
significant deflection and energy dependent time delay in lage
scale extragalactic or halo magnetic fields of strengths
comparable to current upper limits. By performing 
3-dimensional Monte Carlo simulations of nucleon propagation, 
we show how observations of arrival
direction and time distributions can be used to measure the
structure and strength of large-scale magnetic fields, and
constrain the nature of the source of ultra-high energy cosmic
rays.

\end{abstract}

\newpage

\section{Introduction}

If extragalactic magnetic fields (EGMFs) exist, they are at a
level below detectability with presently available techniques.
Faraday rotation measures of distant powerful
radio-sources give upper limits to extragalactic fields of
$B_{\rm rms}\sqrt{l_c}\la 10^{-9}$G\,Mpc$^{1/2}$, where $l_c$
denotes the reversal length of the field  (see, {\it e.g.}, 
Kronberg 1994). However,  EGMFs  below these current limits are
of significant interest to cosmology, galaxy and star formation,
and galactic dynamos (see, {\it e.g.}, Kronberg 1994;
Olinto 1997). 

The propagation of UHECRs is affected by the presence of EGMFs
of strength $10^{-12}\G\la B_{\rm rms}\la10^{-9}\G$ and/or
by Galactic halo fields $10^{-8}\G\la B_H\la10^{-6}\G$.
Protogalactic fields in the former range 
are actually expected if the galactic magnetic field cannot
be explained by a galactic dynamo (Kulsrud \& Anderson 1992).
In this letter, we study how EGMFs and the halo magnetic field,
which we collectively name large-scale magnetic fields (LSMFs),
affect UHECRs with energies $E\ga10\EeV$ (EeV = $10^{18}$ eV).
We simulate the propagation of ultra-high energy nucleons in
the LSMF and the cosmic microwave background (CMB), and show 
how the resulting angle-time-energy images of these
UHECRs, could be used, in turn, to probe LSMFs with strengths in
the above range. Suitable UHECR statistics could be achieved with
future experiments, such as the Japanese Telescope Array 
(Teshima {\it et al.} 1992), the High Resolution
Fly's Eye (Al-Seady {\it et al.} 1996), and the Pierre Auger
Project (Cronin 1992), that have the potential to
detect  $\sim10-10^3$ particles per source with $E\ga10^{19}$eV,
over a $\sim5\yr$ period.
Although the origin and nature of UHECRs are not known,  
they are likely to be generated in extragalactic sources
({\it e.g.}, Bird {\it et al.} 1995; Yoshida {\it et al.} 1995;
Hayashida {\it et al.} 1996), such as powerful radio-galaxies
({\it e.g.}, Rachen \& Biermann 1993), cosmological
$\gamma-$ray bursts (Vietri 1995; Waxman 1995;
Milgrom \& Usov 1996), and/or topological defects
({\it e.g.}, Bhattacharjee, Hill \& Schramm 1992; Sigl 1996).
Our results can also be used to discriminate between models of
the origin of UHECRs.

Previous studies of the effect of the LSMF on UHECRs have
included a discussion of the energy dependent deflection and
time delay for 
extragalactic UHECRs (Cronin 1992; Waxman \&  Miralda-Escud\'e
1996; Medina Tanco {\it et al.} 1997). In addition, the
effect of EGMFs on secondary $\gamma$-rays  produced in the
interaction of ultra-high energy protons with CMB photons can
also probe EGMFs below current upper limits: Plaga (1995) and
Waxman \&  Coppi (1996) proposed to use the arrival time delay
of secondary $\gamma$-rays in the TeV range to probe EGMFs of
strengths $B_{\rm rms}\la10^{-15}$G, while Lee,  Olinto \&
Sigl (1996) showed how fields of  strength 
$B_{\rm rms}\sim10^{-9}$G affect the  $\gamma-$ray spectrum
around $10\EeV$ due to electron synchrotron losses. 

\section{Angle-time-energy images}

Nucleons propagating in intergalactic space with energies
$\ga10\,$EeV are mainly subject to scattering on the LSMFs 
and pair production on the CMB (for protons), as well as
photopion production on the CMB. Pair
production dominates the energy loss below
$\simeq70\,$EeV which we include as a continuous loss process
(Chodorowski, Zdziarski, \& Sikora 1992). Above $\simeq70\,$EeV
photopion production dominates and gives rise  to the
Greisen-Zatsepin-Kuz'min  cutoff (hereafter GZK cutoff; Greisen
1966; Zatsepin \&  Kuzmin 1966). We model photopion production
as a stochastic process where the chance of
interaction is drawn at random, as described in Lee (1996),
and Sigl, Lee, \& Coppi (1996). We model the EGMF as a gaussian 
random field, with zero mean in Fourier space, and  power 
spectrum: $\left\langle B^2(k)\right\rangle\propto k^{n_B}$ for
$k<2\pi/l_c$, and $\left\langle B^2(k)\right\rangle=0$
otherwise. Here, $l_c$ characterizes the cutoff scale of
magnetic fluctuations; we choose $l_c=1\Mpc$ as a fiducial value
for the EGMF. The power spectrum is normalized via:
$B_{\rm rms}^2\equiv V/(2\pi)^3\int d^3{\bf k}B^2({\bf k})$.
We leave the detail description of our Monte-Carlo code to Sigl,
Lemoine, \& Olinto (1997).

A simpler version of the present study was carried out by 
Medina Tanco {\it
et al.} (1997). These authors described the field as an assembly
of randomly oriented bubbles of constant field, with diameter
equal to the coherence length, and neglected the stochastic
nature of pion production.
Finally, they considered the limit $D\theta_E\gg l_c$,
where $D$ denotes the distance to the source, and $\theta_E$ the
r.m.s. deflection angle at energy $E$; this limit is obtained
only for extreme choices of the magnetic field  strength and
configuration for $E\ga10^{20}$eV (see below). The limit
$D\theta_E\ll l_c$ is not only the most probable  for
$E>10^{19}$eV, it also is the most difficult to treat (see
Sigl, Lemoine, \& Olinto (1997).
The importance of distinguishing the limits $D\theta_E\gg l_c$
and $D\theta_E\ll l_c$, was pointed out by Waxman \& 
Miralda-Escud\'{e} (1996). In the latter limit, all nucleons 
emitted by the source and captured on the detector have
essentially experienced the same magnetic structure along
their paths, hence the scatter around the mean of 
the correlations between the time delay $\tau_E$, the deflection
angle $\theta_E$, and the energy $E$, is of order 1\%. In the
opposite limit, $D\theta_E\gg l_c$, the relative widths of
these correlations are of order unity.

Our simulations generate data consisting of the energy, arrival
time, and angular direction for each UHECR emitted from the
source.  For general reference, we can
estimate the average deflection angle $\theta_E$ and  the
average time delay $\tau_E\simeq D\theta_E^2/2c$ induced on a
proton of energy $E\la50\EeV$ over a distance $D$ in the limit
$D\theta_E\gg l_c$. Using (Waxman
\& Miralda-Escud\'{e} 1996): 
\begin{equation}
\tau_E\,\simeq\,2.5\,\left(\frac{3+n_B}{2+n_B}\right)
\left(\frac{D}{10\Mpc}\right)^2
\left(\frac{E}{10\EeV}\right)^{-2} \left(\frac{B_{\rm
rms}}{10^{-11}\G}\right)^2
\left(\frac{l_c}{1\Mpc}\right)\yr\,,
\label{t_delay}
\end{equation}
and $\theta_E\simeq 0.014^\circ(D/10\Mpc)^{-1/2}
(\tau_E/1\yr)^{1/2}$.
The Faraday rotation  bound $B_{\rm
rms}\sqrt{l_c}<10^{-9}$G\,Mpc$^{1/2}$,  implies $\tau_E \lesssim
3\times10^4$ $ (D/10\Mpc)^2 (E/10\EeV)^{-2}\,$yr, and
$\theta_E\lesssim2.5^\circ (D/10\Mpc)^{1/2}$ $(E/10\EeV)^{-1}$,
assuming $n_B\simeq0$. In what follows, we address the possible
observables for different cases. In \S 2.1 and \S 2.2, we
discuss the case of an extra-galactic bursting source. In \S 
2.3, we discuss the case of a continuously emitting source. In 
these sections, we consider the case in which the time delay
induced by the extra-galactic magnetic field dominates over
that of the halo magnetic field of our Galaxy. In \S 2.4, we 
discuss how the domination by a halo field would modify our 
conclusions. We use the notations
$\tau_{100}\equiv\tau_{E=100EeV}$, 
 $\theta_{100}\equiv\theta_{E=100EeV}$.

\subsection{Observable time delays}

The best case for probing large scale magnetic fields as well
as the nature of the source is: 
$\tau_{50}\simeq1\yr$, {\it i.e.}, the time delay is
comparable to the integration time of the experiment, $T_{obs}$.
In this case, the limit  $D\theta_E\ll l_c$ holds as the
Faraday rotation bound is combined with $\tau_{50}\simeq1\yr$. 
For a propagated differential energy spectrum  ${\rm
dJ}\propto E^{-\eta}{\rm d}E$, that corresponds to
the injected spectrum below the GZK cutoff,
Eq.~(\ref{t_delay}) tells us that the arrival time distribution
is given by  ${\rm dJ}\propto t^{(\eta-3)/2}{\rm d}t$, for a
bursting source with emission timescale  $T_S\ll\tau_{50}$ and 
a $\tau_E-E$ correlation with negligible scatter.  Note that a
continuous source with  $T_S\gg1\yr$ would
produce a uniform  distribution of arrival times, notably
independent of their  energy (see \S 2.4).

The observation of the arrival time and energy of two events is 
sufficient to determine a zero-point in time (time of emission), 
hence the value of $\tau_E$, equivalently
$DB_{\rm rms}\sqrt{l_c}$. The distance can be
independently derived by fitting the energy spectrum above the 
GZK cutoff, provided there is enough statistics.
Thus,  one  ends up with an estimate of the distance to the
source, the  nature of the source ({\it e.g.}, burst {\it vs.}
continuous emission), and the value of $B_{\rm rms}^2l_c$.
Observations of several clusters of events would thus
``map'' the extra-galactic magnetic field. Finally, one can
show, from $\tau_{50}\simeq D\theta_{50}^2/2 \simeq 1\yr$,
that the angular image would not be 
resolved for a typical detector with resolution 
$\delta\theta\ga0.5^\circ$. That typical time delays above tens
of EeV may be of the order of a few years, is suggested
by a detailed likelihood analysis (Sigl, Lemoine, \& Olinto 
1997) of the three pairs of events that 
were recently reported by the AGASA experiment (Hayashida 1996).

\subsection{Large time delays}

If $\tau_E\gg T_{obs}\sim{\rm a\,few\,yr}$, a given  
source will be seen only on a limited range in energy
(Waxman \& Miralda-Escud\'{e} 1996); indeed,
protons with higher energy have already reached us in the past,
while protons with lower energy have yet to reach us. These
authors have derived the shape of the energy spectrum in the
limit
$D\theta_E\gg l_c$, for $E\la50$EeV where the signal has
significant scatter $\Delta E/E\sim30$\%; in the opposite limit
$D\theta_E\ll l_c$,  $\Delta E/E\sim1$\%. In principle, the
measure of the width would tell us which limit applies,
and would give us  some information about $B_{\rm
rms}$ and $l_c$. However, this  statement depends strongly on the
distance, as  $D\theta_E/l_c\propto D^{3/2}$.
For sufficiently large time delays, the angular image of the 
source could be resolved for $E\la50\EeV$, in which case the
value of $B_{\rm rms}\sqrt{Dl_c}$ becomes accessible. Moreover,
the argument developed by Waxman \& Miralda-Escud\'{e} (1996) 
would allow to place another constraint on these parameters, 
respectively $D/l_c\ll1$ or $D/l_c\gg1$.

\begin{figure}
\epsscale{0.90}
\plotone{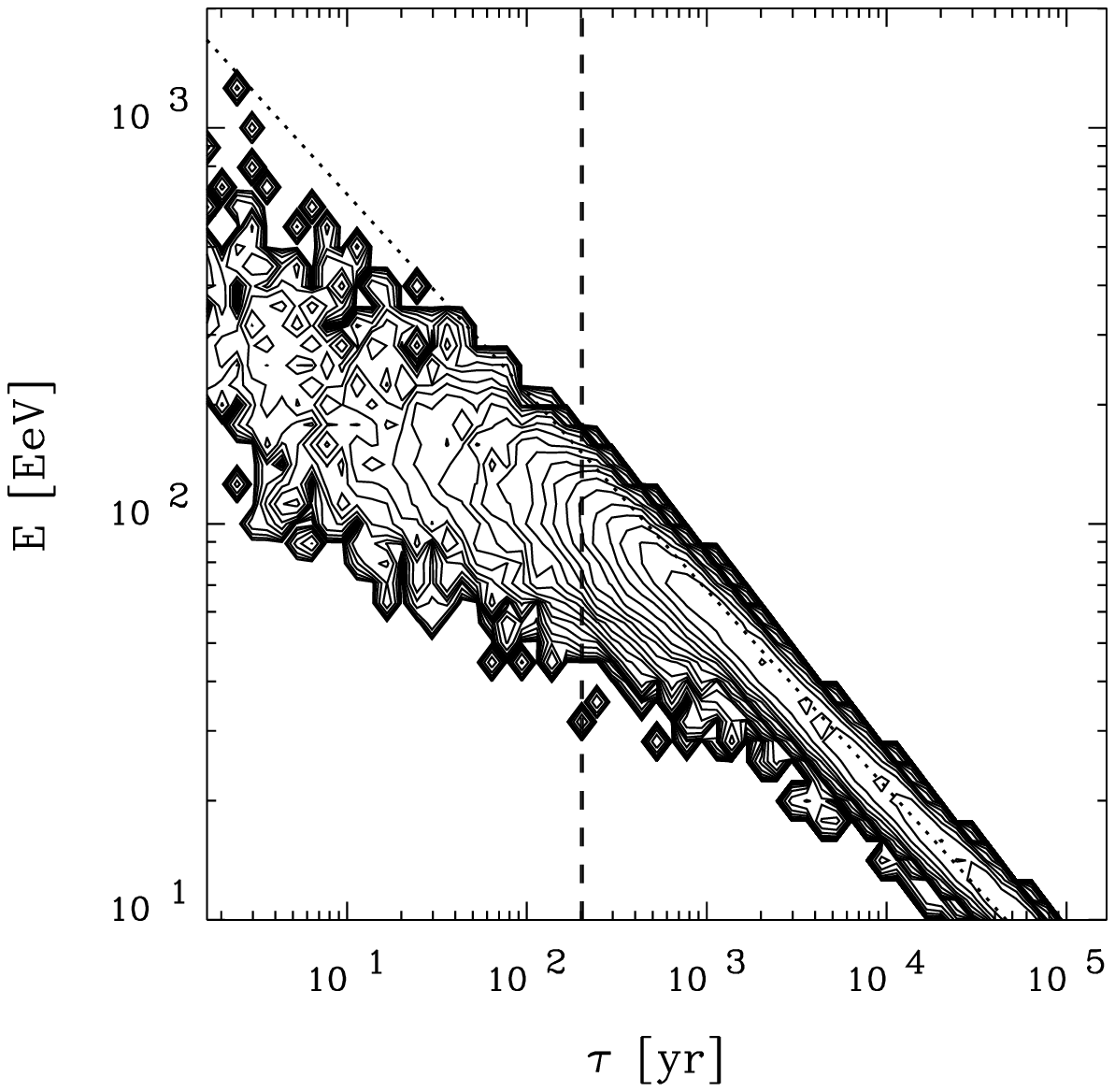}
\caption{Contour plot of the UHECR image projected onto the 
time-energy plane. The distance to the source is 60Mpc;
$B_{rms}=2\times10^{-10}\G$, $l_c=1\Mpc$, $n_B=0$. The 
dotted line indicates the energy-time delay correlation 
$\tau_E\propto E^{-2}$ as would be obtained in the absence of 
pion production losses. The dashed lines, which are not resolved 
here, indicate the location
(arbitrarily chosen) of the observational window, of length
$T_{obs}=5\yr$.}
\label{F1}
\end{figure}

\begin{figure}
\epsscale{0.90}
\plotone{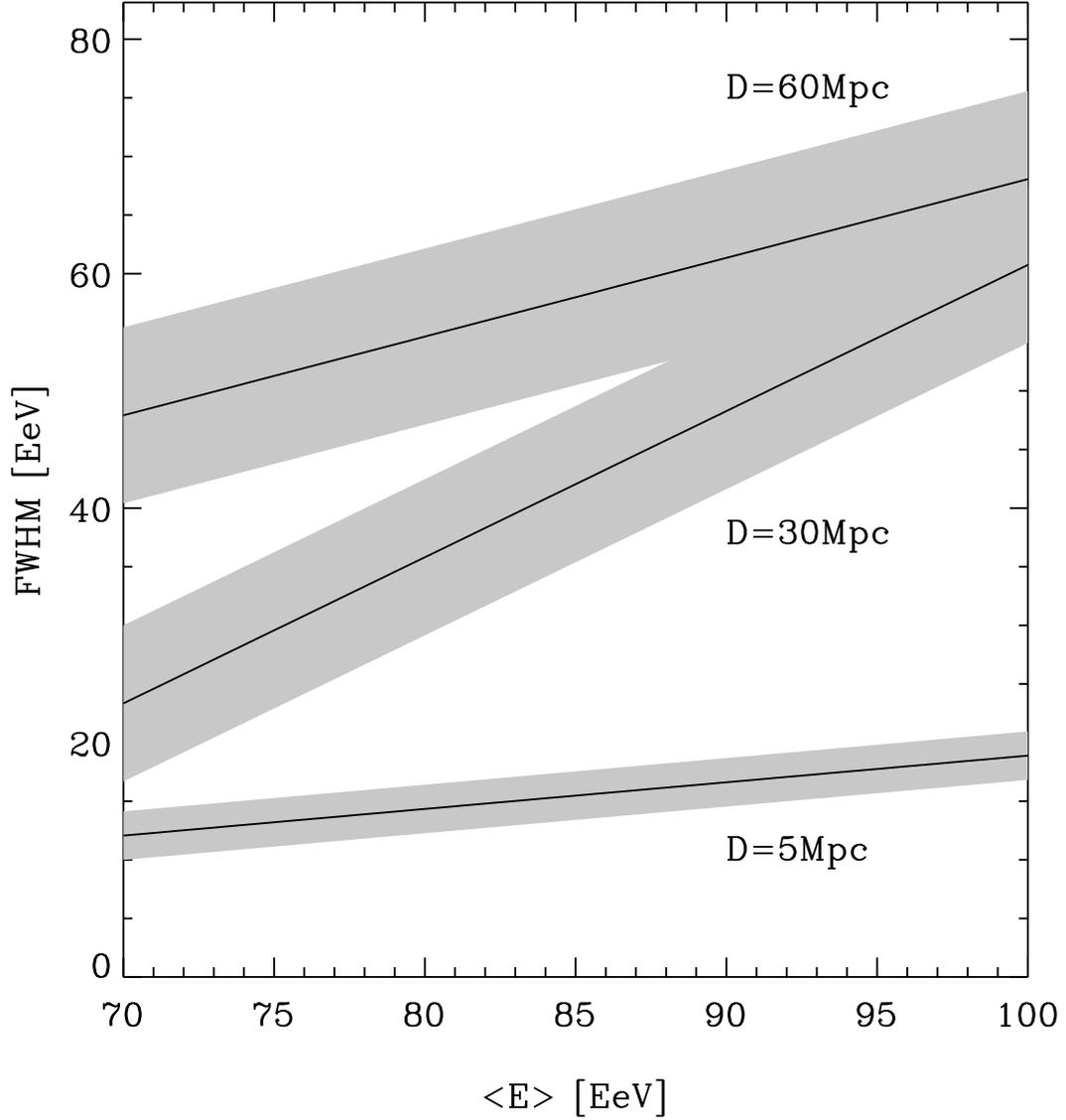}
\caption{Full width at half maximum (FWHM) {\it vs.} the average
energy, for the observed spectrum, above pion production
threshold, for large time delays
$\tau_E\gg T_{obs}$ and small deflection $D\theta_E\ll l_c$. 
Shown are linear fits to the
correlations calculated from time slices through Figure~1
as well as the 1$\sigma$ uncertainty around these fits
for different distances $D=5\Mpc$, $D=30\Mpc$, and $D=60\Mpc$,
as indicated.}
\label{F2}
\end{figure}

For sources that are observed above the pion production
threshold, the previous statements do not apply. However, in
this limit, one can use a similar reasoning to obtain an
estimate of the distance. At a fixed
time, only a given range of energies is observed. Intuitively,
the larger the distance, the more important the pion production,
hence the broader the signal. This effect is illustrated in
Figures~\ref{F1} and~\ref{F2}. Figure~\ref{F1} shows the image
of the source in the time-energy plane for a source lying at
$D=60\Mpc$. The correlation $\tau_E\propto E^{-2}$ is shown
as a dotted line. In this plane, the effect of pion 
production  is mainly to downscatter from higher
energies to lower energies, with a trend toward increasing the
time delay for a given energy at emission.
Figure~\ref{F2} shows the correlation between the width
of the signal in energy, as actually seen by the detector,
{\it vs.} the mean energy of the signal. This signal is
obtained as a slice in the $\tau_E-E$ image, integrated 
between $t$ and $t+T_{obs}$, where $T_{obs}=5\yr$ in this 
case and $t$ is arbitrary; this slice is indicated in
Figure \ref{F1} by the dashed lines. An example of the signal in 
energy so obtained is shown in Figure \ref{F3} in dashed line.
The correlation 
shown in Figure \ref{F2} was obtained by measuring the 
width of the signal for different mean energies, 
corresponding to different choices of $t$, and  adjusting a 
straight line fit. The shaded areas denote the 1$\sigma$ 
range of (numerical) uncertainty. These uncertainties 
actually provide a hint of the actual experimental uncertainties 
associated with such measurements, as the number of particles
used in the Monte-Carlo correspond to that expected from a
typical UHECR source. As Figure~\ref{F2} reveals, an 
estimate of the distance $5\Mpc\la D\la 100\Mpc$, could be
achieved with reasonable accuracy. For $D\la5\Mpc$, the width
of the signal becomes dominated by
the instrumental resolution, whereas for $D\ga100\Mpc$, 
statistics should insufficient.

\subsection{Continously emitting sources}

\begin{figure}
\epsscale{0.90}
\plotone{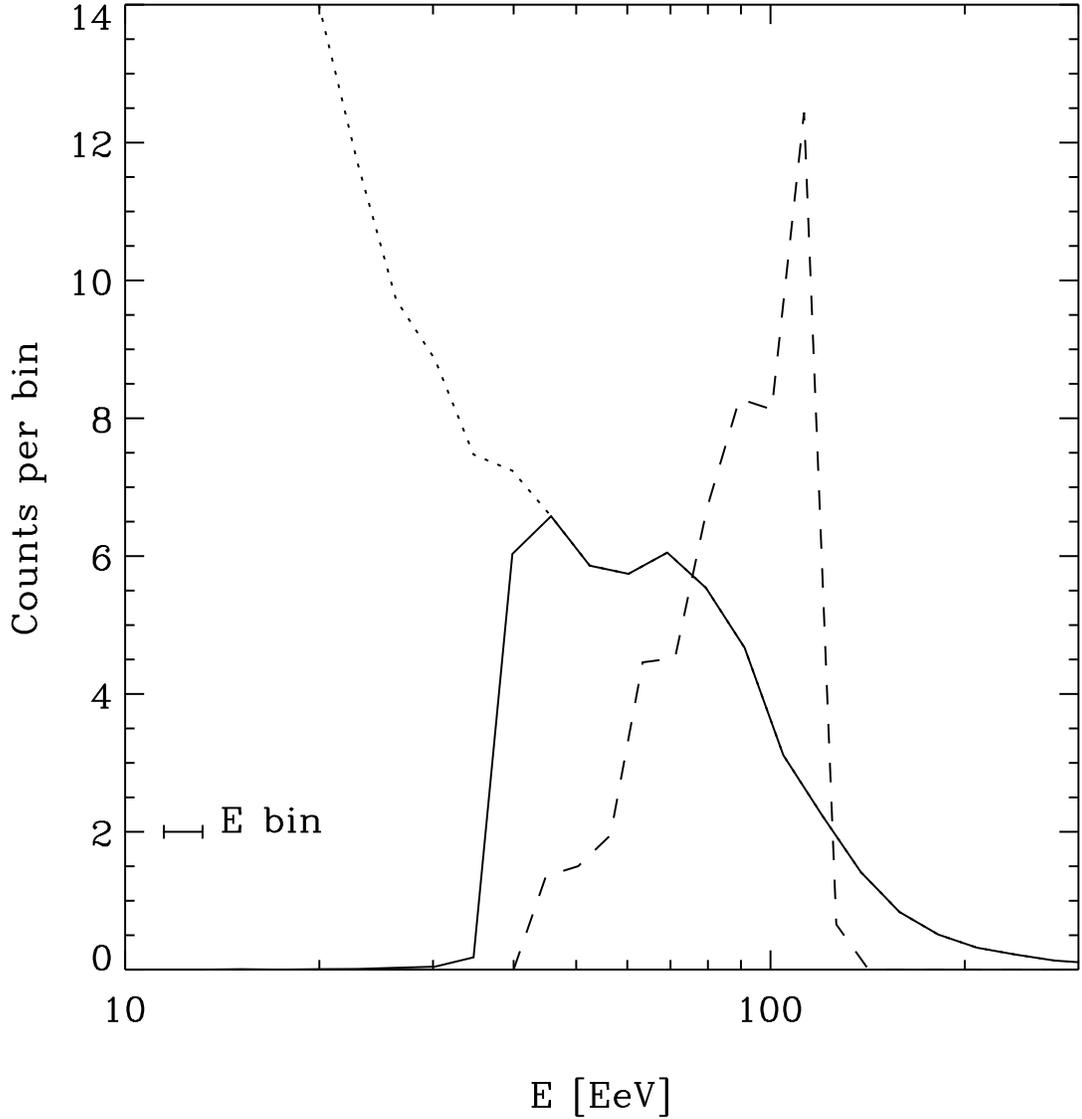}
\caption{Energy spectra for a continuous source (solid 
line), and for a burst (dashed line). Both spectra are 
normalized to a total of 50 particles detected. The parameters 
corresponding to the continuous source case are: $T_S=10^4\yr$, 
$\tau_{100}=1.3\times10^3\yr$, and the time of observation is 
$t=9\times10^3\yr$, relative to propagation with the speed of
light. A low energy cutoff results at the energy $E_S=40\EeV$
where $\tau_{E_S}=t$ (see text). The dotted line shows how the
spectrum would continue if $T_S\ll10^4\yr$. The
case of a bursting source corresponds to a slice of the image in 
the $\tau_E-E$ plane, as indicated in Figure~1 by dashed lines. 
For both spectra, $D=30\Mpc$, and $\gamma=2.$}
\label{F3}
\end{figure}

The case of a continous source, emitting on a timescale
$T_S$, is obtained by folding,
on the time axis, the angle-time-energy image of a
corresponding bursting source, with a top-hat of width $T_S$.
In principle, for a given magnetic field configuration, there
will be an energy $E_S$, such that $T_S=\tau_{E_S}$.
For $E\gg E_S$, no correlation is expected between arrival times
and energies, as the arrival time distribution is dominated by 
the uniform top-hat in this limit. In contrast, for $E\ll E_S$
the source behaves just like a burst with respect 
to the observations; since, in general, $T_S\gg T_{obs}$ (for 
radio-galaxy hot spots, for instance), one also has $\tau_E\gg
T_{obs}$ for $E\la E_S$. Thus,
this situation would  be analogous to that discussed in \S 2.2,
{\it i.e.}, the energy spectrum should show a cut-off
around $\sim E_S$, as the UHECRs with $E\ll E_S$ have not yet 
reached us, even if they were among the first emitted.
The transition between these two regimes is
only observable if $10\EeV\la E_S\la$ a few 100 EeV. Here, the
lower bound is given by the requirement that the deflection
angle in the galactic magnetic field should be less than a few
degrees so that observed events can be associated with a common
source. The upper bound is
dictated by the statistics required to establish reasonable
estimates of the correlations in time and energy. The simulation
of a possible case is shown in Figure~\ref{F3}, where the energy 
spectrum, as seen by the detector, is plotted for a total of 50 
detected particles.

The energy spectrum above $E_S$ can be used to
estimate the distance $D$, for a given injection spectrum, via a
pion production fit (see Figure \ref{F3}). The
angular image should be resolved for sufficiently large time
delays (see \S 2.2), and the actual magnitude of $\tau_E$ can be
derived, as $\tau_E\propto D\theta_E^2$.
Therefore, not only can the magnetic field strength 
$B_{\rm rms}\sqrt{l_c}$ be determined,  the 
timescale of emission is also obtained as a by-product. If
$T_S\sim T_{obs}$, the angular image would 
not be resolved, but the time delay could be directly measured, 
as in \S 2.1, and the above results still hold. In the 
intermediate case, $1\yr\ll T_S\la10^3\yr$, one could only place 
an upper limit on the magnetic field strength,
$B_{\rm rms}\sqrt{l_c}\la10^{-10}\G\,\Mpc^{1/2}
(\theta_E/0.5^\circ)(D/50\Mpc)^{-1}(E/10\EeV)$, hence an upper
limit on $\tau_E$ and $T_S$. For reasonable values of the
distance, this constraint is already more stringent than the
Faraday rotation limit. 

If the field strength
lies close to the Faraday bound, the range
$300(D/30\Mpc)^2\yr\lesssim
T_S\lesssim3\times10^5(D/30\Mpc)^2\yr$ is within reach.
Observation of several sources at various
distances would enlarge the detectable range of emission time
scales and time delays. In case no low-energy cutoff is seen 
down to $\simeq10\EeV$, but the deflection angle is resolved,
the above estimates of $T_S$ turn into lower limits. The
opposite limit is given if $T_S\ll\tau_E$
for all $E$ up to the value where the
experiment runs out of statistics. In this case the source 
behaves like a bursting source and the discussion in \S 2.2
applies. We note that, here, the deflection angle should be
measurable, if $T_S\ga10^4\yr$, hence
$B_{\rm rms}\sqrt{l_c}$ could be measured.

\subsection{Magnetized galactic halo}

In principle, the results of \S 2.1, \S 2.2 and \S 2.3, should be
considered as constraints on both  the EGMF and the halo
magnetic field, in the sense that
$\theta_E^2\,\simeq\,\theta_{XG}^2+\theta_{H}^2$, where 
$\theta_{XG}$ and $\theta_{H}$ denote the 
deflections induced by the EGMF and by the halo field 
respectively.  The halo field has a significant effect if
its strength is in the range $10^{-8}\G\la B_H\la10^{-6}\G$ and
its scale height $\ga 1\kpc$. Present observational constraints
are not conclusive (see, {\it e. g.}, Beck et al. 1996, and 
Kronberg 1994); some authors argue  for significant halo fields
with large scale heights,  while others argue for smaller scale
heights, $\la 1\kpc$.  Therefore,  one cannot {\it a
priori} rule out the possibility that the  deflection and time
delays are dominated by the influence of the  halo magnetic
field, even if the nucleons originate from an  extra-galactic
source.
If the effect of EGMFs are weak, the  results of \S 2.1, \S 2.2, 
and \S 2.3 can be applied to the halo magnetic field by
substituting the extension of the magnetic halo for the distance
to the source. The only exception is that, for a
bursting source, a strong correlation between arrival time and
energy is expected even above the GZK cutoff irrespective of its
distance. This is due to the absence of pion production during
propagation through the halo when most of the time delay and
deflection is accumulated. As well, in case the energy spectrum 
above the GZK cut-off could be observed, for instance for a 
continuous source as in \S 2.3, or a burst with a small time 
delay, the absence of pion production would constitute a 
signature of the proximity of this source.

\section{Conclusions}

We have shown that an EGMF of strength
$10^{-12}\G\la B_{\rm rms}\la10^{-9}\G$, and/or a Galactic halo
magnetic field of strength $10^{-6}\G\la B_H\la 10^{-6}\G$,
leave distinct 
signatures in the angle-time-energy images of UHECRs of energy 
$E\ga10\EeV$, which could be used, inversely, to probe these
fields. Obviously, the respective detection or non-detection
of a correlation between arrival times and energies marks a 
bursting source with a short time delay $\tau_E\sim T_{obs}$, or
a continuous source with emission timescale $T_S\ga T_{obs}$.
A bursting source with a large time delay would be seen with no 
significant correlation between arrival times and energies, 
however, only in a range of energy of width $\Delta E/E\la$30\%
(Waxman \& Miralda-Escud\'e, 1996). 
The observation of an energy 
spectrum (integrated over the observational window), up to the 
highest energies, that shows a cut-off at a low energy (still 
above $\sim10\EeV$) constitutes the signature of a continuous 
source with an activity time scale $T_S$ comparable to the
typical time delay $\tau_E$ at the cut-off energy.
Information on the actual magnitude of $T_S$ is contained in
the high end of the observed spectrum and in the
arrival directions. The absence of a lower cutoff implies
$T_S>\tau_{10\EeV}$.
In the limit of large time delays, deflection angles of events
around $10\EeV$ should be measurable, and the value of
$B_{\rm rms}\sqrt{Dl_c}$ could be derived therefrom. If both the
typical time delay and $T_S$ are smaller than the
integration time, the whole spectrum above $10\EeV$ would be
``scanned through''. In this case, both $D$ and
$B_{\rm rms}\sqrt{l_c}$ could be determined.
A more quantitative implementation of the effects discussed here
should involve a likelihood approach. We have performed such an
analysis for the three UHECR pairs recently suggested by AGASA
(Hayashida {\it et al.} 1996). Although present
data are much to sparse to draw any quantitative conclusions, we
observed some potentially interesting tendencies (Sigl, Lemoine,
\& Olinto 1997). One of the pairs, for instance, turns out to be
inconsistent with a burst and comparatively small time delays of
the order of a few years may be favored. 

Strictly speaking, the analysis in the present paper applies to
models that predict UHECR to be nucleons in the relevant
energy range. Topological defect models predict
a domination of $\gamma-$rays
above $\simeq50\EeV$ ({\it e.g.}, Sigl, Lee, \& Coppi 1996).
However, due to its electronic content, the deflection and delay
of an electromagnetic cascade roughly correspond to those of a 
nucleon of the same energy, modified by the relative lifetime 
fraction of electrons at that energy, typically a factor 
$\approx0.5$ (Lee 1996).
We therefore expect our analysis to reproduce at least the
correct tendencies also for defect models.

\acknowledgments{We acknowledge P. Biermann and A. Dubey for 
useful discussions. We thank the Max-Planck
Institut f\"ur Physik, M\"unchen, Germany and the Institut 
d'Astrophysique de Paris, France for providing CPU time.
We thank the Aspen Center for Physics for hospitality and  
support. G.S. acknowledges financial support by the Deutsche
Forschungs Gemeinschaft under grant SFB 375 and by the
Max-Planck Institut f\"ur Physik. This work was supported, in
part,
by the DoE, NSF, and NASA at the University of Chicago, and
by the DoE and by NASA through grant NAG 5-2788 at Fermilab.}

\clearpage

\end{document}